\documentclass[preprint,3p]{elsarticle} %final form June 23, afternoon ; ends at line 733
\usepackage{amsmath}
\usepackage{amssymb}
\usepackage{caption} 
\usepackage{bigints}
 \usepackage[english]{babel} 
 \usepackage{float} 
 \usepackage{graphics} 
 \usepackage{graphicx} 
\usepackage[utf8]{inputenc}
\usepackage{mathtools}
\usepackage{capt-of} 
\usepackage{calrsfs}
\usepackage{lipsum}  % to provide "filler" text
 \usepackage{epsfig}
\newcommand{\R}{\mathbf{R}}
\newcommand{\C}{\mathbf{C}}

\newtheorem{theorem}{Theorem}[section]
\newtheorem{lemma}{Lemma}[section]

\newdefinition{remark}{Remark}[section]
\newdefinition{fe}{Feature}[section]
\newproof{proof}{Proof}
\newproof{pot}{Proof of Theorem \ref{thm2}}
\newproof{poot}{Proof of Corollary \ref{co1}}

\newcommand{\xx}{\mathbf{x}}

\newcommand{\nn}{\mathbf{n}}

\newcommand{\rr}{\mathbf{r}}
\newcommand{\qq}{\mathbf{q}}  % new command

\newcommand{\FF}{\mathbf{F}}

\newcommand{\RRR}{\mathbf{R}^3}

\DeclareMathAlphabet{\pazocal}{OMS}{zplm}{m}{n}

\numberwithin{equation}{section}
\newdefinition{exe}{Example}[section]
\journal{}
\begin{document}%
\begin{frontmatter}

\title{On generalized and fractional derivatives and their applications to classical mechanics}
\author{Angelo B. Mingarelli\footnote[1]{angelo@math.carleton.ca}}
\address{
$^{1}$School of Mathematics and Statistics, Carleton University, Ottawa, Canada}

\begin{abstract}
A generalized differential operator on the real line is defined by means of a limiting process.  When an additional fractional parameter is introduced, this process leads to a locally defined  fractional derivative. The study of such generalized derivatives includes, as a special case, basic results involving the classical derivative and current research involving  fractional differential operators. All our operators satisfy properties such as the sum, product/quotient rules, and the chain rule.  We study a Sturm-Liouville eigenvalue problem with generalized derivatives and show that the general case is actually a consequence of standard Sturm-Liouville Theory.  As an application of the developments herein we find the general solution of a generalized harmonic oscillator. We also consider the classical problem of a planar motion under a central force and show that the general solution of this problem is still generically an ellipse, and that this result is true independently of the choice of the generalized derivatives being used modulo a time shift. The generic nature of phase plane orbits modulo a time shift is extended to the classical gravitational n-body problem of Newton to show that the global nature of these orbits is independent of the choice of the generalized derivatives being used in defining the force law. Finally, restricting the generalized derivatives to a special class of fractional derivatives, we consider the question of motion under gravity with and without resistance and arrive at a new notion of time that depends on the fractional parameter. The results herein are meant to clarify and extend many known results in the literature and intended to show the limitations and use of generalized derivatives and corresponding fractional derivatives.
\end{abstract}

\end{frontmatter}

\section{Introduction}

Generalized and fractional derivatives have been studied extensively especially in the past few decades. Apparently, the notion of a {\it fractional} derivative is ancient in the sense that such a derivative's existence was questioned by L'Hospital in a letter to Leibniz dated some time in 1695. Since then the notion has been considered by more authors than it is possible to do justice here, in this brief introduction, and so we shall refer the reader to other sources for derivatives and fractional derivatives not considered here. For a brief historical survey of the subject of fractional derivatives  see, for example, the texts \cite{kil}, \cite{pod}, \cite{sam}, and the recent paper \cite{om}, among many others.

The main purpose of this study is to find a formalism under which we can provide a unified proof of the most basic properties of the classical (first) derivative of a real function, i.e., a general derivative concept that would include properties such as the Sum Rule, Product Rule, Chain Rule etc. and then detail some of its applications. The general derivative concept here is approached by means of a limiting process, as is done in the usual case. Thus, for an appropriate real valued function $p$ defined on some subset of the plane, we define the generalized derivative of a given function $f$, defined on the range of $p$, by means of the limit
$$D f(t) = \lim_{h\to 0} \frac{f(p(t,h))-f(t)}{h}$$
whenever the limit exists and is finite (specific assumptions are stated below). Recently, a {\it fractional derivative} of order $\alpha$ (with $0 < \alpha <1$) was defined by means of the limit definition,
$$D^{\alpha} f(t) = \lim_{h\to 0} \frac{f(p(t,h,\alpha))-f(t)}{h}$$
in the special cases where $p(t,h,\alpha) = t+ht^{1-\alpha}$ see \cite{kha}, and where $p(t,h,\alpha) = t\exp(ht^{-\alpha})$ see \cite{kat1}. In both cases the resulting fractional derivative is shown to satisfy the basic rules of differentiation stated above. However, we prefer to refer to these fractional derivatives as {\it local fractional derivatives} because of their radically distinct properties in comparison with those studied in \cite{kil}, \cite{pod}, and \cite{sam}. For the sake of brevity we shall continue refer to derivatives defined by the limiting processes above simply as fractional derivatives.

Under fairly mild conditions on the function $p$ we show in Section 2 that the basic rules of differentiation are satisfied for generalized derivatives defined by $D f(t)$ above, without any reference to the parameter $\alpha$. This shows that the fractional parameter is generally unnecessary if we want a notion of a derivative that extends to the fractional case and, in addition, satisfies the basic properties required of a derivative.  Given that many existing  fractional derivatives {\it do not} satisfy such basic differentiation properties, we are in a unique position to extend the class of fractional derivatives that {\it do} satisfy such properties.

In Section 3, we discuss the generalized derivative so defined and its relationship to the fractional derivatives defined by others, such as in \cite{kat1}, \cite{kha}.

A generalized Sturm-Liouville equation and resulting eigenvalue problem is considered in Section 4 with a view to showing that, under suitable conditions, any regular {\it generalized} (and so including {\it fractional} in the sense described here) Sturm-Liouville boundary value problem with separated homogeneous boundary conditions on a finite closed interval may be reduced to the study of a new but now {\it weighted} Sturm-Liouville boundary value problem for which there is a large amount of existing theory, e.g., \cite{atk}, \cite{atm}, \cite{az}.

In Section 5 we consider various basic physical problems when interpreted with the use of generalized derivatives in lieu of classical derivatives. In the case of motion under a central force we show that the generalized problem admits a solution that is generally an {\it ellipse} in conformity with the classical derivative case and that these ``ellipses" persist independently of the choice of the generalized derivative. It seems to indicate that elliptical motion is pervasive when it comes to central force laws. Even for the n-body problem under an inverse square law (the so called Gravitational n-body problem) all the generalized derivative does, in this respect, is to ultimately allow for a change of the independent temporal variable that keeps the global geometry of solutions intact.

Finally, we reconsider the work of \cite{aeb} and show that in the simple case of fractional motion under gravity without and with resistance, we can make sense of fractional unit measurements by defining a new fractional time that depends on the fractional parameter $\alpha$ that can be identified with {\it cosmic time} as suggested in \cite{aeb}. Technical arguments and proofs are in Section 6.

%%%%%%%%%%%

\section{Main results}

In the sequel $I$ will generally denote a non-empty open interval $I\subseteq \R$, $f: I \to \R$, and $p : U_{\tau} \to \R$ where $U_{\delta}=\{(t,h) : t\in I,  |h|< \delta\}$ for some $\delta >0$ is some generally unspecified neighborhood of $(t,0)$. Unless $\delta$ is needed in a calculation we shall simply assume that this condition is always met. Of course, we always assume that the range of $p$ is contained in $I$. In the sequel, $L(I) \equiv L^1(I)$ is the usual space of  Lebesgue integrable functions on $I$. 

For a given function $p$ of two variables, the symbol $D_p f(t)$ defined by the limit
\begin{equation}\label{da}
D_p f(t) = \lim_{h\to 0} \frac{f(p(t,h))-f(t)}{h},
\end{equation}
whenever the limit exists and is finite, will be called the {\bf $\mathbf p$-derivative of $f$} at $t$ or the generalized derivative of $f$ at $t$ and, for brevity, we also say that $f$ is {\boldmath $p$}{\bf -differentiable} at $t$. In the event that $I$ is a closed interval we define the $p$-derivative at the end-points as being the respective one-sided derivatives as is customary. As usual, $f$ is said to be {\it differentiable} whenever its ordinary derivative exists. It is clear from the definition that for any constant $c$, $D_p (c\,f) (t) = c \cdot D_p f(t) .$ Higher order derivatives are defined by composition.

{\bf Notation:} \quad For a fixed function $p$, when there is no risk of confusion, its $p$-derivative at $t$ will be denoted simply by $D f(t)$. Whenever we are dealing with different functions $p$, $q$, \ldots, the various generalized derivatives will be denoted by $D_p^{\alpha} f(t)$,  $D_q^{\alpha} f(t)$, as the case may be. The ordinary derivative of $f$ will be denoted by $f^{\prime}(t)$. Whenever it exists, the partial derivative of $p$ with respect to $h$ will be denoted by $p_h(t,h)$, or just $p_h$.

\begin{exe}\label{ex1a}
We show that there exists a function $p$ such that the $p$-derivative of every differentiable function $f$ on $I$ is identically zero. 

Let  $p(t,h) = t+h^2$ and $I $ be any interval that contains the point $0$ in its interior. Let $f^{\prime}(t)$ exist for all $t\in I$. Then, for all $t\in I$,
$$D_p f(t) = \lim_{h\to 0} \frac{f(t+h^2) - f(t)}{h} = \lim_{h\to 0} \frac{f(t+h^2) - f(t)}{h^2}\cdot h = f^{\prime}(t)\cdot 0 = 0.$$ 
\end{exe} 
Next, one can presume that mere $p$-differentiability necessarily implies continuity, but this is false, in general. We show now that there exists a function $f$, whose $p$-derivative exists at $t=0$ for some appropriate $p$, but yet $f$ fails to be continuous at $0$.

\begin{exe}\label{ex2a}
Let $I $ be any interval that contains the point $0$ in its interior, $p(t,h) = t+h^2$, $f(t) = {\rm sgn}\,(t)$, $f(0)=1$. Then $f$ is $p$-differentiable at $t=0$ yet $f$ is not continuous there. This is clear, since
$$D_p f(0) = \lim_{h\to 0} \frac{f(h^2)-f(0)}{h} = 0.$$
However, $f$ is not continuous at $t=0$. 
\end{exe}

\begin{remark}
Observe, however, that in Example~\ref{ex2a}, $f$ is right-continuous at $t=0$. Also, note that if we use $f(0)= -1$ in lieu of $f(0)=1$, then $f$ is not $p$-differentiable at $t=0$. 
\end{remark}

{\bf The main hypotheses}

In various parts of this paper we shall be referring to one or more of the following assumptions.
\begin{itemize}
\item[${\rm H1}^+$]   For $t\in I$ and for all sufficiently small $\varepsilon > 0 $, the equation $p(t,h) = t+ \varepsilon$ has a solution $h = h(t, \varepsilon)$. In addition, $h \to 0$ as $\varepsilon \to 0$.

\item[${\rm H1}^-$]   For $t\in I$ and for all sufficiently small  $\varepsilon > 0 $, the equation $p(t,h) = t - \varepsilon$ has a solution $h = h(t, \varepsilon)$. In addition, $h \to 0$ as $\varepsilon \to 0$.

%%%%%%% check continuity in Theorem 2.1.ok.

\item[${\rm H2}$]   $\displaystyle \frac{1}{p_h(\cdot,0)}  \in L(I)$.
\end{itemize}

%%%%%%%%%%%%%%%
\begin{lemma}\label{le1}
Assume ${\rm H1}^+$ holds at $t=a$ and that $p(t,h)$ is continuous at $h=0$. If $f$ is $p$-differentiable at $t=a$ then $f$ is right-continuous at $a$.
\end{lemma}
\begin{remark}\label{rem21}
It is difficult to weaken the first assumption in Lemma~\ref{le1}  further without losing right-continuity. It is possible that $p(t,h) = t+\varepsilon$ has no solutions $h$ whatsoever for any  $\varepsilon>0$. To see this, let $I=(-1,1)$, $p(t,h)=t$, $f(t) = {\rm sgn}\,(t)$ for $t\neq 0$ and $f(0)= - 1$. Even though $D_p f(0)=0$ we observe that $f$ is not right-continuous at $t=0$.

On the other hand, the mere existence of a solution $h(t,\varepsilon)$ in ${\rm H1}^+$ is insufficient to guarantee continuity. For example, let $p(t,h) = t+e^h$, for $t \in I=(-1,1)$, $f(t) = {\rm sgn}\,(t)$ for $t\neq 0$, and $f(0)= 1$. In this case $h =\log \varepsilon $ but now $h$ violates the limit condition in H1.  In this case $f$ is $p$-differentiable at $t=0$ but $f$ is not continuous there. 
\end{remark}

\begin{lemma}\label{le2}
Assume ${\rm H1}^-$ holds at $t=a$ and that $p(t,h)$ is continuous at $h=0$. If $f$ is $p$-differentiable at $t=a$ then $f$ is left-continuous at $a$.
\end{lemma}
Combining Lemma~\ref{le1} and Lemma~\ref{le2} we get,
\begin{theorem}\label{th1}
Let $p$ satisfy both ${\rm H1}^{\pm}$ and that $p(t,h)$ is continuous at $h=0$. If $f$ is $p$-differentiable at $a$ then $f$ is continuous at $a$.
\end{theorem}

\begin{remark}
This key property of $p$-differentiability implying continuity and the properties that follow have been shown in the special cases where $p(t,h) = t+ht^{1-\alpha}$ when $t>0$, $0 < \alpha\leq 1$, see \cite{kha}, and where $p(t,h) = t\exp(ht^{-\alpha})$ when $t>0$, see \cite{kat1}, and $0 < \alpha\leq 1$. It is readily verified that both functions $p$ defined there satisfy ${\rm H1}^{\pm}$, except at $t=0$ which is excluded from $I$, a point where both $p(0,h)= 0$ independently of $h$. Indeed, in \cite{kat1}, \cite{kha}, $f$ need not be continuous at $t=0$ even though it is $p$-differentiable there. Our results include the cited ones as a special case.  As one gathers from the proofs herein, the presence of the fractional parameter $\alpha$ is generally unnecessary so that the results below include all the analogous fractional results mentioned.
\end{remark}

 \begin{theorem} \label{th2} 
\begin{itemize} 
\item[(a)] {\rm (The Sum Rule)}  If $f, g$ are both $p$-differentiable at $t\in I$ then so is their sum, $f + g$, and
$$D (f+ g)(t) =  D f(t) + D g(t).$$
\item[(b)] {\rm (The Product Rule)}  Assume that $p$ satisfies ${\rm H1}^{\pm}$ and that for $t\in I$, $p(t,h)$ is continuous at $h=0$. If $f, g$ are both $p$-differentiable at $t\in I$ then so is their product, $f\cdot g$, and
$$D (f\cdot g)(t) =  f(t)\cdot D g(t) + g(t)\cdot D f(t). $$
\item[(c)] {\rm (The Quotient Rule)}  Assume that $p$ satisfies ${\rm H1}^{\pm}$ and that for $t\in I$, $p(t,h)$ is continuous at $h=0$. If $f, g$ are both $p$-differentiable at $t\in I$ and $ g(t) \neq 0$ then so is their quotient, $f/g$, and
$$D \left (\frac{f}{g}\right)(t) =  \frac{g(t)\cdot D f(t) - f(t)\cdot D g(t)}{g(t)^2}. $$
\end{itemize}
\end{theorem}

Whenever the fractional parameter {\it does} appear explicitly in the definition of $p$, we replace $D$ above and below by $D^{\alpha}$ and obtain the results in those papers (recall that $0 < \alpha<1$). See the next section for more details on such fractional derivatives.

\begin{theorem}\label{th3} {\rm (The Chain Rule)}
Assume that $p$ satisfies ${\rm H1}^{\pm}$ and that for $t\in I$, $p(t,h)$ is continuous at $h=0$. Let $f$ be continuous and non-constant on $I$, and let $f$ be $p$-differentiable at $t\in I$. Let $g$ be defined on the range of $f$ and be differentiable at $f(t)$. Then the composition $g\circ f $ is $p$-differentiable at $t$ and
\begin{equation}\label{eq0}
D (g\circ f )(t) = g^{\prime}(f(t))\,D f(t).
\end{equation}
\end{theorem}
The relationship between differentiability and $p$-differentiability is next and shows that the {\it fractional derivatives} defined in \cite{kat1} and \cite{kha} are simply first order weighted classical derivatives, when the latter exist (see Remark~\ref{rem32} below).
\begin{theorem}\label{th4}
Let $p$ satisfy ${\rm H1}^{\pm}$ and for  $t\in I$ let $p(t, h)$, $p_h(t,h)$ be continuous in a neighborhood of $h=0$ with   $p_h(t,0) \neq 0$. Finally, let $\lim_{\varepsilon\to 0} \varepsilon/h(t,\varepsilon)$ exist and be non-zero.  Then $f$ is differentiable at $t$ iff and only if  $f$ is $p$-differentiable at $t$. In addition, whenever $f$ is differentiable at $t$, then 
\begin{equation}\label{eq5}
D_p f(t) = p_h(t,0)\, f^{\prime}(t).
\end{equation}
\end{theorem}

\begin{remark}
The assumption $p_h(t,0) \neq 0$ cannot be waived, in general, \emph{i.e.,} the result is false if $p_h(t,0) =0$. In other words,  there exist functions $p, f$ such that $f$ is $p$-differentiable at $t$ but not differentiable there. To see this let $p(t,h) = t+h^3$, $f(t) =  |t|$, $I$ be any interval containing $0$ in its interior. Then $p_h(t,0) =0$ for all $t$.  However, a glance shows that $D_p f(0) =0$ yet $f^{\prime}(0)$ does not exist as a classical first order derivative. The ``$\varepsilon/h$" condition is readily verified for each of the basic derivatives considered in \cite{kat1}, \cite{kha}.

%It now follows that there exists a function $f$, namely \emph{ Weierstrass' continuous and nowhere differentiable function}, such that $D_p f(t)$ does not exist anywhere on $\R$ for any of the functions $p$ under consideration. %Of interest here is the determination of a function $p$ satisfying ${\rm H1}^{\pm}$ and satisfying the other hypotheses of Theorem~\ref{th4}, except that now we require $p_h(t,0)=0$, and such that $D_p f(t)$ is nowhere $p$-differentiable on $\R$. 
\end{remark}

\section{Fractional derivatives}

We now introduce the notion of a {\bf fractional derivative of order $\mathbf \alpha$}. This is defined by \eqref{da} once again except that $p$ is now a function of the three variables $(t,h,\alpha)$, $(t,h)\in U$, and $\alpha$ is a parameter with $0 < \alpha <1$, usually, but this restriction is not necessary, in general, see e.g., \cite{kat1}, \cite{kha}. Thus, for a given $\alpha$, we define the fractional derivative of order $\alpha$ by 
\begin{equation}\label{da1}
D_p^{\alpha} f(t) = \lim_{h\to 0} \frac{f(p(t,h,\alpha))-f(t)}{h},
\end{equation}
where it is understood that $\alpha=1$ gives the ordinary derivative. As before, we will denote $D_p^{\alpha} f(t)$ simply by $D^{\alpha} f(t)$ when there is no confusion. It is shown in \cite{kat1}, \cite{kha} that the Product/Quotient/Chain Rule is valid for such fractional derivatives in the special cases where $p(t,h,\alpha) = t+ht^{1-\alpha}$ and  $p(t,h,\alpha) = t\exp(ht^{-\alpha})$ but these results are now clear, as our proofs  are  independent of $\alpha$. Thus, any definition of a fractional derivative of order $\alpha$ defined by means of a limiting process such as \eqref{da} \emph{must} satisfy the basic rules of ordinary differentiation and is essentially a parameter-dependent weighted first order ordinary differential operator.
\begin{remark}
One may anticipate here that, in lieu of \eqref{eq0}, $D^{\alpha} (g\circ f )(t) = (D^{\alpha} g)(f(t))\,D^{\alpha} f(t)$, hold for $g$ being merely $\alpha$-differentiable at $f(t)$. However, this too   is generally false as we show presently, see [\cite{kha}, p. 66] in this regard. Choose $I = [0,1]$, $g(t) = t$, $f(t)=t$, $p(t,h,\alpha) = t+ht^{1-\alpha}$ where $0 < \alpha < 1$. Then, an easy calculation shows that $D^\alpha (g\circ f)(t) = t^{1-\alpha}$ for all $t\in I$, $(D^\alpha g)(f(t))=t^{1-\alpha}$, and  $D^\alpha g(t) =t^{1-\alpha}$ for any such $\alpha$ and any $t \in [0,1]$. Thus, unless $\alpha=1$ which is clear, $$D^\alpha (g\circ f )(t)  \neq (D^{\alpha}g) (f(t))\, D^\alpha f(t),\quad t\in (0,1)$$
in general. 

\begin{exe}
We recall that the Khalil-Horani-Yousef-Sababhehb fractional derivative is defined by setting $p(t,h,\alpha) = t+ ht^{1-\alpha}$ where $0<\alpha<1$, \cite{kha}. The Katugampola fractional derivative is obtained by using $p(t,h,\alpha) = te^{h\,t^{-\alpha}}$ where $0<\alpha<1$, \cite{kat1}. It is known that both the Katugampola and Khalil fractional derivatives coincide for $0<\alpha<1$ when $f$ is differentiable, (see [\cite{kat1}, Theorem 2.3\, (7)] and [\cite{kha}, Theorem 2.2\, (6) ]).
\end{exe}

Of course, if $f(t) \equiv C$ where $C$ is a constant then $D^{\alpha} f(t) \equiv 0$. On the other hand, we have seen in Example~\ref{ex1a} that there are {\it non-identically  constant} functions $f$ such that $D f(t) = 0$  a.e. An $\alpha$-explicit example of this type may be created using $p(t,h,\alpha) = t + h^2|t|^{1-\alpha}$, $t>0$,  with a minor variation of the argument in said example.

\end{remark}

\begin{remark}\label{rem32}
The result corresponding to Theorem~\ref{th4} for $\alpha$-fractional derivatives is \emph{a fortiori} true and can be stated as
$$D_p^{\alpha} f(t) = p_h(t,0,\alpha)\, f^{\prime}(t),$$
whenever $f$ is differentiable at $t$.
\end{remark}

\begin{remark}
It follows from Theorem~\ref{th4} that such generalized derivatives (including $\alpha$-fractional derivatives as defined here) are no more than an $\alpha$-dependent multiplication operator on $V$, where $V$ is the derived set (or the set of all derivatives) of the space of all differentiable functions.

Note that there are {\it no sign restrictions} on either the function $p$ nor its partial derivative $p_h(t,0)$. This is important for later applications, especially to Sturm-Liouville theory, as we shall see next.

%From the preceding theory and remarks it follows that, when it comes to a theory of $p$-differentiability, \emph{the case $p_h(t,0)=0$ is worthy of investigation, more so than the case $p_h(t,0)\neq 0$.}
\end{remark}

\section{Sturm-Liouville Theory}

We will deal with two main applications, the first is to formulate a  Sturm-Liouville operator using fractional derivatives and show that a study of its spectrum is, in fact, equivalent to the study of the spectrum of a {\it weighted} Sturm-Liouville eigenvalue problem. In the two-term case,
\begin{equation}\label{eq6}
-D^2 y = - D \circ D y = \lambda\, y, \quad\quad t\in I
\end{equation}
we can find the general solution exactly in terms of circular trigonometric functions.  In the case where $I$ is a finite closed interval, $I=[a,b]$, we append real separated homogeneous boundary conditions of the form
\begin{gather}
y(a)\,\cos \mu - D y(a)\, \sin \mu=0, \label{eq7}\\
y(b)\,\cos \nu + D y(b)\, \sin \nu =0, \label{eq8}
\end{gather}
to obtain a self-adjoint eigenvalue problem of a traditional type but with a possibly sign indefinite weight function and a possibly indefinite principal part (which in itself may destroy the Hilbert space self-adjointness of the operators involved in some cases).

%\subsection{Sturm-Liouville theory}

We assume that $p$ satisfies H2 and \eqref{eq5} in addition to the standing hypotheses ${\rm H1}^{\pm}$ and that $I$ is a closed bounded interval $[a,b]$. In the sequel all integrals are understood in the Lebesgue sense so we take it that solutions to \eqref{eq6} are at least absolutely continuous, the space of such being denoted by $AC(I)$. 

Fix $\lambda \in \R$ and some function, $p$. A {\bf solution} of \eqref{eq6} is a function $y \in AC(I)$ such that $D y \in AC(I)$ and $y(t)$ satisfies \eqref{eq6} a.e. on $I$. Since $y^\prime$ exists a.e. it follows by Theorem~\ref{th4} that $D y(t)$ exists a.e. and $D y(t) = p_h(t,0)\, y^\prime(t)$ a.e. on $I$. On the other hand, since $D y \in AC(I)$, $(D y)^\prime$ exists a.e. and so $D y$ is also $p$-differentiable (a.e.) with 
\begin{equation}\label{eq9a}
-D^2\, y(t)  = - p_h(t,0)\,(p_h(t,0)\, y^\prime(t))^\prime, \,\, {\rm a.e.}
\end{equation} 
Suppressing the ``{\rm a.e.}" proviso for simplicity, we get that 
any solution of \eqref{eq6} satisfies the ordinary differential equation
\begin{equation}\label{eq10}
- (p_h(t,0)\, y^\prime(t))^\prime = \lambda\, \frac{1}{p_h(t,0)}\, y(t).
\end{equation}
Such solutions are usually called {\it Carath\'eodory} solutions of \eqref{eq10} and are now standard nomenclature in the study of Sturm-Liouville equations with measurable coefficients, see, for example, \cite{er} for one of the earliest such references.

Indeed, under H2 the general solution of \eqref{eq10} for $\lambda \neq 0$ is easily seen to be given by
\begin{equation}\label{eq11}
y(t) = A\,\sin \left (\sqrt{\lambda}\, \int_a^t \frac{1}{p_h(s,0)}\, ds \right ) + B \, \cos \left (\sqrt{\lambda}\, \int_a^t \frac{1}{p_h(s,0)}\, ds \right ),
\end{equation}
where $A, B$ are constants and we fix the determination of the square root if  $\lambda <0$. 
\begin{remark}
It is important to note that the boundary value problem \eqref{eq6}, \eqref{eq7}-\eqref{eq8} is governed by a boundary  condition on the $p$-derivative of $y$, not the ordinary derivative of $y$.
\end{remark}
\begin{remark}
Observe that, on account of \eqref{eq9a},\eqref{eq10} and \eqref{eq11}, we can now readily study problems of the form $D^2 y(t) =\lambda\, y(t)+ f(t)$ using standard methods (variation of parameters, etc.) and so no new  theory is necessary.
\end{remark}
We now refer the reader to standard texts on Sturm-Liouville Theory such as \cite{atk}, \cite{atm}, \cite{az} and the references therein for proofs of the statements contained in the following. Given that the study of the fractional/generalized Sturm-Liouville equations considered here reduces to the study of ordinary weighted Sturm-Liouville equations we can state the following result as a sample of what can be achieved.

\begin{theorem}\label{th5}
Let $I=[a,b]$. Assume that $p_h(t,0) > 0$ a.e. for $t\in I$ and $q/p_h \in L^1(I)$, $w/p_h \in L^1(I)$. Finally let
$$\int_a^b \sqrt{\left ( \frac{w(s)}{p_h(s,0)}\right )_{\pm}}\, ds > 0,$$
\end{theorem}
where the symbols $\pm$ appended to the function indicate the positive and negative part of the function (i.e., $f(t)_{\pm} = \max \{\pm f(t), 0\}$). Then the generalized Sturm-Liouville eigenvalue problem
\begin{equation}\label{eq13}
-D^2 y + q(t)y = \lambda\, w(t)\, y 
\end{equation}
subject to a pair of real separated homogeneous boundary conditions, \eqref{eq7}-\eqref{eq8},
%\begin{gather}
%y(a)\,\cos \mu - D y(a)\, \sin \mu=0, \label{eq14}\\
%y(b)\,\cos \nu + D y(b)\, \sin \nu =0, \label{eq15}
%\end{gather}
where $\mu, \nu \in [0, \pi)$, has a doubly infinite sequence of real eigenvalues $\lambda_n^{\pm}$ where $\lambda_n^{\pm} \to \pm \infty$ as $n \to \infty$ having the asymptotic distribution
\begin{equation*}
\displaystyle \lambda_n^{\pm} \sim \pm\, \frac{n^2\pi^2}{\left ( \bigintss_a^b \sqrt { ( \frac{w(s)}{p_h(s,0)})}_{\pm}\, ds \right )^2},
\end{equation*}
 as $n \to \infty$, \cite{atm2}.

There is also an at most finite (and possibly empty) set of non-real eigenvalues occurring in complex conjugate pairs, the total number of which never exceeds the number $N$, of negative eigenvalues of the problem 
\begin{equation}\label{eq16}
- (p_h(t,0)\, y^{\prime})^{\prime} + \frac{q(t)}{p_h(t,0)}\, y = \lambda\, y 
\end{equation}
subject to the same boundary conditions \eqref{eq7}-\eqref{eq8}.  The total number of non-simple real eigenvalues is also finite and bounded by the same constant, $N$, above.

\begin{remark}
The assumption $p_h >0$ in Theorem~\ref{th5} above cannot be relaxed, in general. For example, consider the case where $p(t,h,\alpha) = t+{\rm sgn}\,(t)\,h$, $I=(-1,1)$, and  \eqref{eq6}-\eqref{eq8} with $\mu=0, \nu=0$. Then \eqref{eq6} with generalized derivative defined in \eqref{da} leads to the {\it degenerate} Sturm-Liouville operator 
$$ - {\rm sgn}\,(t)({\rm sgn}\,(t)\,y^{\prime})^{\prime} = \lambda\,y \Longleftrightarrow - ({\rm sgn}\,(t)\,y^{\prime})^{\prime} = \lambda\, {\rm sgn}\,(t)\,y$$
whose spectrum consists only of eigenvalues which fill {\bf all} the complex plane, $\C$, see \cite{atm2}.
\end{remark}
\begin{remark}
In the case of Theorem~\ref{th5} there is also an {\it oscillation theorem} for the eigenfunctions due to Haupt-Richardson, which is a little complicated to state, so we refer the interested reader to the survey paper \cite{abm}. The point here is that the study of the spectrum of the fractional problem reduces to the study of the spectrum of a traditional weighted problem of Sturm-Liouville type.
\end{remark}
\begin{remark}
The authors [\cite{alr}, Remark 3] are correct in suggesting, but not proving, that in the special case where $p(t,h,\alpha) = t + ht^{1-\alpha}$, $0< \alpha<1$, $I=(0,1)$, the problem \eqref{eq13}-\eqref{eq7}-\eqref{eq8} has an infinite number of real eigenvalues, as this is now a simple consequence of the theory presented here and standard Sturm-Liouville theory in the regular case, \cite{az}. 
\end{remark}
Now consider the more general Sturm-Liouville equation
\begin{equation}\label{eq26}
-D\left ( P(t)\,D y\right ) + Q(t)y = \lambda\, W(t)\, y,\quad t\in I=[a,b].
\end{equation}
Within the framework described above, this is then equivalent to the problem
 \begin{equation}\label{eq26a}
- p_h(t,0)\,(p_h(t,0)\,P(t)\, y^\prime(t))^\prime + Q(t)y = \lambda\, W(t)\, y,\quad t\in I,
\end{equation}
with boundary conditions
\begin{gather}
y(a)\,\cos \mu - (P(\cdot)\,p_h(\cdot,0)\,y^{\prime})\,(a)\, \sin \mu=0, \label{eq26b}\\
y(b)\,\cos \nu + (P(\cdot)\,p_h(\cdot,0)\,y^{\prime})\,(b)\, \sin \nu =0. \label{eq26c}
\end{gather}
In order for the problem \eqref{eq26a}-\eqref{eq26c} to be well defined as a regular problem, we impose the assumptions,
$$ \frac{1}{P(\cdot)\,p_h(\cdot,0)}, \quad \frac{Q(\cdot)}{p_h(\cdot,0)},\quad \frac{W(\cdot)}{p_h(\cdot,0)} \in L(I).$$
As a technicality, we will also assume that there is no set of positive measure on which $1/p_h(\cdot,0) = 0$ (or else we need to appeal to the Atkinson reformulation  of Sturm-Liouville theory as a first order system [\cite{atk}, p. 202], which we shall not do here for the sake of clarity and brevity). 

In this case, the change of independent variable
\begin{equation}\label{eq27}
t \to \tau = \int_a^t \frac{ds}{P(s)\,p_h(s,0)}
\end{equation}
$y(t)=z(\tau)$, although {\it not generally invertible} since there are no sign conditions on either $P, p_h$ and so the map $t \to \tau$ is not generally one-to-one, maps \eqref{eq26} into the equation in $\tau$ given by 
\begin{equation}\label{eq28}
-z^{\prime\prime}+ Q^*(\tau)\,z = \lambda W^*(\tau)\,z, \quad \tau \in J
\end{equation}
where $Q^*(\tau) = Q(\tau)\,P(\tau)$, $W^*(\tau) = W(\tau)\,P(\tau)$. In addition, $I$ is mapped into an interval $J$ (but not in a one-to-one fashion)  defined by $J = [0, c]$ where $$c=\max_{t\in [a,b]}\int_a^t \frac{ds}{P(s)\,p_h(s,0)}.$$
Depending on the sign properties of $P(t)\,p_h(t,0)$ for $t\in I$, a number of inverse transformations may still be defined locally.

On the other hand, if $P(t)\,p_h(t,0) $ is a.e. of one sign on $I$ then a global inverse transformation for \eqref{eq27} is well defined and \eqref{eq28} falls within the framework of Sturm-Liouville theory \cite{az}, so that \eqref{eq26}  is really an ordinary regular weighted Sturm-Liouville equation in disguise.

\section{Generalized and fractional mechanics}

Applying the preceding theory to mechanics would require \emph{time reversibility} at the very least. Indeed, Newton's Law,
\begin{equation}\label{eq22n}
F = m \frac{d^2x}{dt^2}
\end{equation}
with classical derivatives is time reversible since replacing ``$t$" by ``$-t$" leaves the equation invariant and this is a necessary condition for doing classical mechanics in our universe. This is most easily accomplished in \eqref{eq9a} by requiring that our function $p$ satisfy a time symmetry assumption such as
$$p_h(t,0) = p_h(-t,0)$$
for all $t$. (Indeed, $p(t,h) = t+h$ works in the case of classical derivatives.) Another good choice for doing physics (with time reversibility) is the $\alpha$-dependent (fractional) choice,
\begin{equation}\label{eq22a}
p(t,h,\alpha) = t+|t|^{1-\alpha}\, h,\quad\quad t\in [-1,1],\quad 0 < \alpha < 1,
\end{equation}
considered in \cite{alr}, \cite{chu}, \cite{aeb}, \cite{kha} but {\bf only} in the case where $t>0$, an unnecessary assumption therein. Clearly $p$ as defined in \eqref{eq22a} satisfies both ${\rm H1}^{\pm}$  and H2 whenever $0<\alpha<1$ as assumed in \cite{alr},\cite{kha}.

In this case the operator on the left of \eqref{eq6} satisfies %i.e., $ D^{2} $ where now $D \equiv D_p^{\alpha}\equiv D^{\alpha}$ acting on absolutely continuous functions $f$ on $[-1,1]$ for which  $|t|^{1-\alpha}\, f^{\prime}(t)$ is also absolutely continuous is time reversible since, by Theorem~\ref{th4}  and \eqref{eq9a}, 
$$D^{\alpha}\circ D^{\alpha}f = |t|^{1-\alpha}\, (|t|^{1-\alpha}\, f^{\prime}(t))^{\prime},$$
%which, as a differential operator, clearly satisfies 
and the time reversibility condition. %In any case, we showed earlier that, under suitable conditions, such $p_h(t,0)$-terms may be transformed out of the equations by means of a time-variable change leading to new differential equations with classical derivatives.

\subsection{Motion under a central force}

In this section we show that doing physics with generalized derivatives as defined here leads to results compatible with doing physics using classical derivatives. This is interesting in the sense that it seems to imply that regardless of {\it how} we define the notion of a derivative, so long as it is of the form \eqref{da}, we still get meaningful results as far as some general laws are concerned. We assume throughout that all generalized derivatives are induced by functions $p$ satisfying ${\rm H1}^{\pm}$ and H2.

Recall that if $\FF = \nn\, F(r)$ is a {\it central force} in the plane where $\nn = \rr/r$, then in  rectangular coordinates its components are given by
$$F_1 = \frac{x}{r}\,F(r),\quad F_2 = \frac{y}{r}\,F(r),$$
where $r^2=x^2+y^2$. The equations of motion of a particle subjected to such a force using {\it generalized derivatives} are then given by 
$$m\,D^2\,x = \frac{x}{r}\,F(r),\quad\quad m\,D^2\, y = \frac{y}{r}\,F(r).$$
Now consider, for example, a particle with mass $m$ attracted to a fixed point $O$ by a central force $F(r) = -m\,k^2\,r$ where $k$ is a constant. As before, in ordinary rectangular coordinates the equations of motion using generalized derivatives now become
\begin{equation}\label{eq18}
m\,D^2\,x = - m\,k^2\,x,\quad\quad  m\,D^2\, y = -m\,k^2\,y,
\end{equation}
or, using Theorem~\ref{th4}, and simplifying we find
\begin{gather}
- (p_h(t,0)\, x^\prime(t))^\prime = \lambda\, \frac{1}{p_h(t,0)}\, x(t), \label{eq19} \\
- (p_h(t,0,)\, y^\prime(t))^\prime = \lambda\, \frac{1}{p_h(t,0)}\, y(t)\label{eq20},
\end{gather}
where $\lambda=k^2 >0$. If we take it that $[a,b]=[0,1]$, say, then both of these equations  \eqref{eq19}-\eqref{eq20} have the general solution (see \eqref{eq11}) ,
\begin{gather}
x(t) = c_1\,\sin \left (\sqrt{\lambda}\, \int_0^t \frac{1}{p_h(s,0)}\, ds \right ) + c_2 \, \cos \left (\sqrt{\lambda}\, \int_0^t \frac{1}{p_h(s,0)}\, ds \right ), \label{eq21} \\
y(t) = d_1\,\sin \left (\sqrt{\lambda}\, \int_0^t \frac{1}{p_h(s,0)}\, ds \right ) + d_2 \, \cos \left (\sqrt{\lambda}\, \int_0^t \frac{1}{p_h(s,0)}\, ds \right ), \label{eq22}
\end{gather}
{\it regardless} of any sign conditions on $p_h$! From these we easily derive the relation
\begin{equation}\label{eq22x}
(d_1x-c_1y)^2+(d_2x-c_2y)^2 = (d_1c_2-c_1d_2)^2,
\end{equation}
where (recall that $(p_h\,x^\prime)(t),(p_h\,y^\prime)(t) \in AC(I)$),
$$c_1 = \frac{1}{\sqrt{\lambda}}\,(p_h\,x^\prime)(0),\quad c_2 = x(0),$$
$$d_1 = \frac{1}{\sqrt{\lambda}}\,(p_h\,y^\prime)(0),\quad d_2 = y(0).$$
Observe that the phase plane motion $(x(t),y(t))$ defined by the relation \eqref{eq22x} as $t \in [0,1]$ varies in any way possible,  must lie on an {\it ellipse} whose center is at the center of the force for all $t\in [0,1]$. This is completely analogous to the case of classical derivatives. In fact, this shows that {\it the ellipticity conclusion is independent of the definition of the generalized derivative} being used to define the laws of motion (subject only to our Definition~\ref{da} and a temporal shift). In other words, the geometry or qualitative picture of this generalized equation of motion \eqref{eq18} is the same as the original (classical derivative) motion modulo a generally nonlinear time scale shift. Thus, in \eqref{eq18},  particles still move on ellipses but with a different time scale (see \eqref{eq27} for the shifted time defined by $\tau$).

%\begin{exe}\label{ex3}
%The solution of the (generalized) initial value problem \eqref{eq18},
%$$(x(0),y(0))= (c,0),\quad\quad (D x(0), Dy(0))= (0,v_0)$$
%consisting of a particle starting at $(c,0)$ with a (generalized) initial velocity $(0, v_0)$ in the vertical direction is given by (recall \eqref{eq5}) 
%$$x(t) = c\, \cos \left (\sqrt{\lambda}\, \int_0^t \frac{1}{p_h(s,0)}\, ds \right ),\quad\quad y(t) = \frac{v_0}{\sqrt{\lambda}}\, \sin \left (\sqrt{\lambda}\, \int_0^t \frac{1}{p_h(s,0)}\, ds \right ).$$
%\end{exe}
%\begin{remark}
%In general we cannot replace the generalized derivative in the initial condition in Example~\ref{ex3} above by a classical derivative without possibly losing the absolute continuity of some of the solutions involved.
%\end{remark}

\subsection{The gravitational n-body problem}
In this section we apply the preceding ideas to show that the phase portraits of the classical Newtonian n-body problem is essentially independent of the choice of generalized derivatives used, modulo a possibly significant time shift.

The classical equations of motion of n (point mass) particles of mass $m_i>0$, $i=1,2,\ldots,n$, with positions $\rr_i(t)={\rm col}\,(r_{i1}(t),r_{i2}(t),r_{i3}\,(t)) \in \RRR$, $i=1,2,\ldots,n$, under an inverse square law are given by
\begin{gather}
m_i\, D^2\rr_i =  \sum_{j=1, j\neq i}^n\frac{G\,m_i\,m_j}{|\rr_j-\rr_i|^3}\,(\rr_j-\rr_i)\label{eq400}
\end{gather}
where $G$ is the gravitational constant and $D^2= d^2/dt^2$. Replacing the classical derivatives by generalized derivatives as per \eqref{da} and using the theory developed here we get an  analogous system 
%%%%%%%%% stopped here, 10:16 p.m. Feb 23.
\begin{gather}\label{eq40}
m_i\, p_h(t,0)\,( p_h(t,0)\, \rr_{i}^{\prime})^\prime  =  \sum_{j=1, j\neq i}^n\frac{G\,m_i\,m_j}{|\rr_j-\rr_i|^3}\,(\rr_j-\rr_i)
\end{gather}
Assuming an integrability condition such as $1/p_h(\cdot,0) \in L(I)$, where $I$ is some unspecified but fixed interval, and say $p_h(t,0) > 0$ a.e. on I, then the time change of variable $t\to \tau$ defined by \eqref{eq27} with $P(t)\equiv 1$, $r_{ij}(t) = q_{ij}(\tau)$, reduces  \eqref{eq40} to a set of equations of the form
\begin{gather}
 m_i\, \frac{d^2 \qq_i}{d\tau^2} =  \sum_{j=1, j\neq i}^n\frac{G\,m_i\,m_j}{|\qq_j-\qq_i|^3}\,(\qq_j-\qq_i)\label{eq43}
\end{gather}
where the inverse of the transformation \eqref{eq27} is now well-defined (as $p_h(t,0) > 0$). Of course, this is the same set of equations as \eqref{eq400} but over a different time scale. It follows that there is a one-to-one correspondence between the global phase plane geometry of the orbits of either system \eqref{eq400} or \eqref{eq43} and that, once again, the phase plane geometry (but not the time frame) is independent of the choice of generalized derivatives. 

This method can be applied to other second order equations in mathematical physics with appropriate restrictions to show that passing to generalized derivatives preserves global geometry of the general solution in the classical derivative case modulo a time shift of the form \eqref{eq27} with $P=1$. It is interesting to determine which other physical laws are {\it generic} in this sense. %i.e., that the result is, in some sense, generically independent of the choice of the generalized derivative being used in question. 
Certainly, autonomous systems of the form $\xx^{\prime\prime} = f(\xx)$, $\xx \in \R^n$, can be considered for which a similar theory arises.

\subsection{Motion under gravity}

Here we consider the basic question of motion under gravity of a particle of mass $m$, without resistance of any kind. This problem was considered recently in \cite{aeb} for a $p$-term of the form \eqref{eq22a} with $t>0$ and in \cite{fa} for the {\it Caputo fractional derivative}, see \cite{kil},  \cite{pod}, \cite{sam}. (However, Caputo derivatives are not of the form treated in this theory.) 

In the usual classical derivative case this consists of a set of two equations of the form
$$m\,x^{\prime\prime} = 0, \quad m\,y^{\prime\prime} = - m g,$$
where $g$ is the acceleration due to gravity, and the derivatives are time derivatives. Its general solution is, of course,
\begin{equation}\label{eq23}
x(t) = x_{_0}+u_{_0}\,t,\quad y(t) = y_{_0} + v_{_0}\,t- \frac12 g\, t^2,
\end{equation}
where $x_{_0}, u_{_0}, y_{_0}, v_{_0}$ are initial given conditions on the position and initial velocities in each direction. Thus, the motion is parabolic in the $(x,y)$-plane, as is basic in mechanics.

Now consider Newton's Law \eqref{eq22n} in terms of generalized derivatives ($m\neq 0$, of course). The corresponding question then leads to a set of two equations of the form
$$m\,D^2\,x =0,\quad\quad m\,D^2\, y = -m\,g,$$
which is equivalent to (see Theorem~\ref{th4}),
$$ p_h(t,0)\,(p_h(t,0)\, x^\prime(t))^\prime =0,\quad\quad p_h(t,0)\,(p_h(t,0)\, y^\prime(t))^\prime = -\,g,$$
whose general solution is readily given by
\begin{eqnarray*}
x(t) &=&  x_{_0}+u_{_0}\,\int_0^t \frac{1}{p_h(s,0)} \, ds,\\
y(t) &= & y_{_0} + v_{_0}\, \int_0^t \frac{1}{p_h(s,0)}\, ds - g\, \int_0^t \frac{1}{p_h(s,0)}\, \left ( \int_0^s \frac{1}{p_h(s_1,0)}\, ds_1\right ) \, ds.
\end{eqnarray*}

The special case $p(t,h,\alpha) = t+|t|^{1-\alpha}\, h$ introduced earlier for $0 < \alpha<1$ but now restricted to $I=[0,1]$ is of some interest, \cite{aeb}. In this case, the latter become
\begin{gather}\label{eq24}
x(t) =  x_{_0}+u_{_0}\, \frac{t^\alpha}{\alpha},\\
y(t) = y_{_0} + v_{_0}\, \frac{t^\alpha}{\alpha} - \frac{1}{2\alpha^2}\, g\, t^{2\alpha}.\label{eq25}
\end{gather}
The two equations \eqref{eq24}-\eqref{eq25} when compared with \eqref{eq23} reveal a \emph{slow time} in the sense that the particle is somewhat higher than it should be at the same time, so it's as if time had slowed down. It is evident that  \eqref{eq24}-\eqref{eq25} reduce to the usual case (with ordinary derivatives) as $\alpha \to 1^-$.

\begin{remark}\label{rem51}
{\it Dimensional Analysis} of \eqref{eq24}-\eqref{eq25}. Recall that $u_{_0} = D^{\alpha}x(0) = (t^{1-\alpha}x^{\prime})(t)$ evaluated at $t=0$. Hence the units of the initial {\it velocity} $u_{_0}$ are not m/$\sec$ as expected in the MKS/ISU-system, but actually $\sec^{1-\alpha}\cdot\, {\rm m}/\sec$ or m$/\sec^{\alpha}$. This then gives the correct units in \eqref{eq24}, namely, m, the unit of distance. Thus, velocities such as $u_{_0}, v_{_0}$ in \eqref{eq24}-\eqref{eq25} have the units m$/\sec^{\alpha}$. The question now arises, ``How does one measure such generalized velocities in terms of this {\it fractional} or $\alpha$-time?"

Recall that a generalized velocity here is of the form given by Theorem~\ref{th4} with $f(t)$ being the distance traveled at time $t$ and $p$ is as defined in this section. Thus,
$$D^{\alpha}f(t) = t^{1-\alpha}\, \frac{df}{dt},$$
so that, in order to get the correct units, we must have that the quantity $t^{\alpha-1}\, D^{\alpha}f(t)$ have units equal to  m$/\sec$. Since the units for generalized velocities are  m$/\sec^{\alpha}$, for $0 < \alpha <1$, we introduce another unit of time, called simply an $\mathbf \alpha${\bf -second}, denoted by $\sec^{\alpha}$, and defined by,
\begin{equation}\label{eq25a}
{\rm 1\,\, sec} = (9192631770)^{1 -\alpha}\,\, \sec^{\alpha},
\end{equation}
where $9192631770$ refers to the {\it period equal to 9,192,631,770 cycles of the radiation, which corresponds to the transition between two energy levels of the ground state of the Cesium-133 atom} which, since 1967, has been accepted as the ISU standard unit for the second, \cite{isu}. Using this definition we can calculate generalized velocities and accelerations so that all units are consistent in \eqref{eq24}-\eqref{eq25}. Recall that $\alpha \in (0,1)$ in this section.  For example, if $\alpha = 0.99$, use of \eqref{eq25a} shows that a velocity of
\begin{eqnarray*}
{\rm 3\, m}\,\, /\sec &=&  3\cdot (9192631770)^{\alpha-1}\,\, {\rm m}/\sec^{\alpha}\\
&=& 3\cdot (9192631770)^{-0.1}\,\, {\rm m}/\sec^{0.99}\\
&=&  2.38\,\, {\rm m}/\sec^{0.99}
\end{eqnarray*}
Clearly, the closer $\alpha$ is to 1, the closer the generalized velocities are to ordinary/classical velocities.

Now consider a generalized  {\it acceleration} defined by the composition of two such generalized derivatives, 
$$D^{2\alpha}\,f(t) =t^{1-\alpha}\, (t^{1-\alpha}f^{\prime})^{\prime}(t),$$
acting on absolutely continuous functions. The nested term, a generalized velocity, has units m$/\sec^{\alpha}$ and so its derivative gives the new term, $(t^{1-\alpha}f^{\prime})^{\prime}(t)$ with units of the form m$/\sec^{\alpha+1}$. Multiplying this by a term with units $\sec^{1-\alpha}$ gives the generalized acceleration units equal to m$/\sec^{2\alpha} $. Thus, we need to change the numerical value of $g$ in \eqref{eq25} in order to compensate for this new system of  {\it fractional time} units.
So, for example, if 
$\alpha = 0.99$, use of \eqref{eq25a} again shows that $g$, the  acceleration due to gravity,
\begin{eqnarray*}
{\rm 9.8\, m}\,\, /\sec^2 &=&  9.8\cdot (9192631770)^{2\alpha-2}\,\, {\rm m}/\sec^{2\alpha}\\
&=& 9.8\cdot (9192631770)^{-0.2}\,\, {\rm m}/\sec^{1.98}\\
&=&  6.19\,\, {\rm m}/\sec^{1.98}
\end{eqnarray*}
Once the numerical value of $g$ is modified, the term $gt^{2\alpha}$ in \eqref{eq25} has units of distance as required, as do all the other terms. As before, the closer $\alpha$ is to 1, the closer the generalized accelerations are to ordinary/classical accelerations.

In \cite{aeb} the authors introduced an unknown {\it cosmic time} factor, $\sigma$, having the dimensions of seconds, into the definition of the fractional time derivative induced by the choice \eqref{eq22a} with $t>0$. This led to the correspondence
$$\frac{d}{dt} \to \frac{1}{\sigma^{1-\alpha}}\, \frac{d^\alpha}{dt^\alpha} = \frac{t^{1-\alpha}}{\sigma^{1-\alpha}}\,\frac{df}{dt}.$$
Comparing our results from \eqref{eq25a} with the previous display from \cite{aeb} we see that we are proposing that $\sigma = 9192631770$.
\end{remark}

\subsection{Motion under gravity with resistance}

In this final subsection we consider the problem of the motion of an object (such as a rocket) through vertical re-entry in an  atmosphere subject to air resistance that varies as the square of the velocity, see \cite{nasa}. Thus, taking the positive direction as downward, let
\begin{equation}\label{res00}
m\,v^\prime (t) = m\, g - \frac{C\,\rho\, A}{2}\, v(t)^2
\end{equation}
be the classical ordinary differential equation modeling a body of mass $m$ traveling with velocity $v(t)$,  through a medium of   density $\rho$. Here $C$ is the drag coefficient, $A$ is the cross-sectional area and $g$ is the acceleration due to gravity. We show here that the {\it slow time} effect observed above can be used to model air and similar resistance questions without the need for additional terms in the equations of motion in the classical setting, but just by modifying $\alpha$ and/or $p$.

For example, for a given $p(t, h, \alpha)$ and some real $\alpha$ such that $1/p_h(t,0, \alpha) \in L^1(0,\infty)$, 
 the generalized differential equation
\begin{equation}\label{res02}
 m\,D_p^\alpha v(t) := m\,p_h(t, 0,\alpha) \, v^\prime (t)  =  m\,g,
\end{equation}
has a solution $v(t)$ (generally depending on $\alpha$) given by
\begin{equation}\label{res03}
v(t) = \sqrt{\frac{2mg}{C\rho A}}\, - g\,\int_{t}^{\infty} \frac{ds}{p_h(s,0,\alpha)},
\end{equation} satisfying the condition
$$\lim_{t\to\infty} v(t) =  \sqrt{\frac{2mg}{C\rho A}}.$$

In particular, for $0 < \alpha < 1$, one can show directly that the function $v(t,\alpha)$ defined by 
\begin{equation}\label{res031}v(t,\alpha) = \sqrt{\frac{2mg}{C\rho A}}\, \tanh(c\,(\alpha + t^\alpha)),
\end{equation}
where $c$ is the constant $$c = \frac{\sqrt{2}\,\sigma^{1-\alpha}}{2\,\alpha\, m}\, \sqrt{g\,C\,\rho\, A\, m},$$ chosen so that $v(t, \alpha)$ converges to some solution $v(t)$ of \eqref{res00} as $\alpha \to 1$, satisfies the fractional differential equation
\eqref{res02}
with $p_h(t,0,\alpha) = t^{1-\alpha}\, \sigma^{\alpha-1}\cosh^2(c\,(\alpha + t^\alpha))$.

\begin{remark}
Observe that the terminal velocity, $v_{\rm term}$, is the limit of $v(t)$ as $t\to \infty$ in either \eqref{res03} or \eqref{res031}. So, 
$$v_{\rm term} = \sqrt{\frac{2mg}{C\rho A}},$$
this quantity being {\it identical} to the terminal velocity of the classical equation obtained by setting the right side of \eqref{res00} equal to zero. So, there are infinitely many possibilities in the choice of $p_h(t, 0,\alpha)$ that could be used to model the problem \eqref{res00} and still yield the correct terminal velocity, independently of $\alpha$.

In addition, note that in the absence of a generalized derivative in  \eqref{res02} we just get a classical problem {\it without any  resistance} whatsoever. Thus, generalized derivatives can be used to model problems with resistance once $p$ is chosen appropriately, something that cannot be done in \eqref{res02} if $D^\alpha(t) = v^\prime(t)$. 
\end{remark}

\section{Proofs}

\begin{proof}(Lemma~\ref{le1})
We start by noting that ${\rm H1}^{+}$ and the continuity assumption $p(t,h) \to p(t,0)$ as $h\to 0$ both imply that if $h$ is a solution of $p(t, h) = t + \varepsilon$ we can pass to the limit as $\varepsilon \to 0$ to see that $p(t,0)=t$ for all $t\in I$.
Given $t=a$, let $\varepsilon_n > 0$ be a fixed yet otherwise arbitrary sequence such that $\varepsilon_n \to 0$. By ${\rm H1}^+$ there is a sequence $h_n \in \R$ (depending on $t, \varepsilon$ ) such that $p(a,h_n) = a + \varepsilon_n$ and $h_n \to 0$ as $n \to \infty$.  If an infinite number of $h_n=0$, we have  $f(a + \varepsilon_n) - f(a) = f(p(a,h_n))-f(a) = 0$ and since $p(a,0)=a$, the left side of \eqref{eq00} is automatically satisfied. So, without loss of generality we may assume that all $h_n \neq 0$. Then
\begin{equation}\label{eq00}
f(a+\varepsilon_n)-f(a) = \frac{f(p(a,h_n))-f(a)}{h_n}\, h_n \to Df(a)\cdot 0 = 0,
\end{equation}
as $n\to \infty$, since $f$ is $p$-differentiable at $a$. Repeating this argument for all other such sequences $\varepsilon_n$,  it follows that $f$ is right-continuous at $t=a$. 
\end{proof}%

\begin{proof}(Lemma~\ref{le2})
The proof parallels the preceding one (i.e., Lemma~\ref{le1}) with the necessary changes and so we omit it.
\end{proof}%

\begin{proof}(Theorem~\ref{th1}) 
This is now an immediate consequence of the previous two lemmas, since $f$ is now both left- and right-continuous at $t=a$.
\end{proof}

\begin{proof}(Theorem~\ref{th2})  (a) follows from the usual properties of limits. 

In order to prove (b), we start by noting that the assumptions  imply that $p(t,0)=t$ (see the proof of Lemma~\ref{le1}, above). 

Next, write $k(t) = f(t)\,g(t)$. For $ h \neq 0$, 
\begin{eqnarray*}
\frac{k(p(t,h))-k(t)}{h} &=& f(p(t,h))\, \frac{g(p(t,h))-g(t)}{h}+ g(t)\, \frac{f(p(t,h))-f(t)}{h}.
\end{eqnarray*}
Since $f$ is $p$-differentiable, Theorem~\ref{th1} now implies that, as $h \to 0$, $f(p(t,h)) \to f(p(t,0)) = f(t)$. The result follows.

The proof of (c) is similar. Writing $k=f/g$ we see that
\begin{eqnarray*}
\frac{k(p(t,h))-k(t)}{h} &=& \frac{1}{g(p(t,h))\,g(t)}\, \left ( g(t)\, \frac{f(p(t,h))-f(t)}{h}- f(t)\, \frac{g(p(t,h))-g(t)}{h} \right ).
\end{eqnarray*}
Passing to the limit as $h \to 0$ and arguing as in (b) the result follows.
\end{proof} 

\begin{proof}(Theorem~\ref{th3}). 
The assumptions on $p$ imply that, as $h \to 0$, $p(t,h) \to p(t,0) = t$ (see the proof of Lemma~\ref{le1}). Since $f$ is continuous, $f(p(t,h)) \to f(t)$ also, as $h \to 0$. On the other hand, since $g$ is differentiable at $f(t)$, it is continuous there, so that, as $h \to 0$, $g(f(p(t,h))) \to g(f(t))$. Next, the differentiability of $g$ at $f(t)$ along with the preceding remarks imply that
\begin{equation}\label{eq1}
g(f(p(t,h))) - g(f(t)) = (f(p(t,h)) - f(t))\, \left (g^{\prime}(f(t))+\varepsilon_1 \right)
\end{equation}
where $\varepsilon_1 \to 0$ as $h \to 0$. Furthermore, since $f$ is $p$-differentiable at $t$ we have
\begin{equation}\label{eq2}
f(p(t,h))) - f(t) = h\, \left (D f(t)+\varepsilon_2 \right)
\end{equation}
where $\varepsilon_2 \to 0$ as $h \to 0$. 

Write k(t) = g(f(t)). Combining \eqref{eq1} and \eqref{eq2} we get
\begin{eqnarray*}
k(p(t,h)) - k(t) &= & g(f(p(t,h))) - g(f(t)) \\
&=& (f(p(t,h)) - f(t))\, \left (g^{\prime}(f(t))+\varepsilon_1 \right)\\
&=& h\, \left (D f(t)+\varepsilon_2 \right)\, \left (g^{\prime}(f(t))+\varepsilon_1 \right).
\end{eqnarray*}
Dividing by $h$ and taking the limit as $h \to 0$ gives the conclusion.
\end{proof}

\begin{proof}(Theorem~\ref{th4})
Fix $t$ and let $h > 0$ be sufficiently small. Then
$$\frac{p(t,h) - p(t,0)}{h} = p_h(t,\xi),\quad {\rm where}\,\, 0 < \xi < h.$$ 
Since $p(t,0) = t$ (see the proof of Lemma~\ref{le1}) the previous display yields
\begin{eqnarray*}
\frac{f(p(t,h))-f(t)}{h} &=& \frac{f(t +p_h(t,\xi)\, h)-f(t)}{h}\\
&=& \frac{f(t +\varepsilon)-f(t)}{\varepsilon}\, p_h(t,\xi) \\
\end{eqnarray*}
where $\varepsilon \equiv p_h(t,\xi)\, h$.  Let $h \to 0$. Then $\xi\to 0$ and so $p_h(t,\xi)\to p_h(t,0) \neq 0$.  Hence $\varepsilon \to 0$ so using the differentiability of $f$ at $t$ yields the right $p$-differentiability of $f$ and also \eqref{eq5}. A similar argument is used to establish left $p$-differentiability of $f$. Thus $f$ is $p$-differentiable at $t$.

Conversely, let $f$ be $p$-differentiable at $t$. Since
\begin{eqnarray*}
\frac{f(p(t,h))-f(t)}{h}&=&   \frac{f(t +\varepsilon)-f(t)}{\varepsilon} \cdot \frac{\varepsilon}{h}
\end{eqnarray*}
Letting $\varepsilon \to 0$ (and so $h \to 0$ also)  we see that $f$ is necessarily differentiable at $t$.

\end{proof}

\begin{proof}(Theorem~\ref{th5})
This follows the ideas laid out in the introduction to this section. Since we are looking for absolutely continuous solutions, \eqref{eq9a} shows that, for fixed $\alpha$, \eqref{eq13} is identical to
\begin{equation}\label{eq17}
- (P(t)\, y^\prime(t))^\prime + Q(t)\, y = \lambda\, W(t)\, y
\end{equation}
where $P(t) =p_h(t,0,\alpha)>0$,  $Q(t) = q(t)/p_h(t,0,\alpha)$ and $W(t) = w(t)/p_h(t,0,\alpha)$ and both $Q, W$ satisfy absolute integrability conditions on $I$. This is now a ``regular" Sturm-Liouville problem and since there are no sign conditions on either $q(t)$ or $w(t)$, it is in the {\it non-definite case} (so generally not self-adjoint), see \cite{abm}. The proof of the first statement on the existence of the doubly infinite sequence of real eigenvalues and their asymptotic behavior is now clear because of the results in \cite{atm2} for \eqref{eq17}. The statements dealing with the non-real spectrum and non-simple eigenvalues may be found in \cite{abm} and the references therein.
\end{proof}
\section{Acknowledgment}
The author wishes to thank Chiara Mingarelli (Flatiron Institute) for some helpful comments and the referees for their thorough reading of the manuscript.
%==========================================================================

\end{document}